\documentclass[11pt,a4paper]{article}
\usepackage{amsmath,latexsym,fullpage,amssymb,color}
\usepackage{ifthen,graphics,epsfig}
\usepackage[english]{babel}
\usepackage{times}
\usepackage{float}
\usepackage{xspace}
\usepackage[T1]{fontenc}
\newlength {\squarewidth}

\newtheorem{theorem}{Theorem}
\newtheorem{lemma}{Lemma}

\newcommand{\toto}{xxx}
\newenvironment{proofT}{\noindent{\bf Proof }}
{\hspace*{\fill}$\Box_{Theorem~\ref{\toto}}$\par\vspace{3mm}}
\newenvironment{proofL}{\noindent{\bf Proof }}
{\hspace*{\fill}$\Box_{Lemma~\ref{\toto}}$\par\vspace{3mm}}

\newenvironment{lemma-repeat}[1]{\begin{trivlist}
\item[\hspace{\labelsep}{\bf\noindent Lemma~\ref{#1} }]}%
{\end{trivlist}}

\newenvironment{theorem-repeat}[1]{\begin{trivlist}
\item[\hspace{\labelsep}{\bf\noindent Theorem~\ref{#1} }]}%
{\end{trivlist}}

\newcounter{linecounter}
\newcommand{\linenumbering}{\ifthenelse{\value{linecounter}<10}
{(\arabic{linecounter})}{(\arabic{linecounter})}}
\renewcommand{\line}[1]{\refstepcounter{linecounter}\label{#1}\linenumbering}
\newcommand{\resetline}[1]{\setcounter{linecounter}{0}#1}
\renewcommand{\thelinecounter}{\ifnum \value{linecounter} > 
9 \else \fi\arabic{linecounter}}
\newfloat{algorithm}{thp}{lop}
\floatname{algorithm}{Algorithm}
\newfloat{construction}{thp}{lop}
\floatname{construction}{Construction}
\newcommand{\Xomit}[1]{}
\newcommand{\REG}{\mathit{REG}}
\newcommand{\CAMP}{\cal CAMP}

\begin{document}

\title{\bf  Two-Bit Messages are Sufficient to Implement\\
            Atomic Read/Write Registers in Crash-prone  Systems}

\author{Achour Most\'efaoui$^{\dag}$, 
        Michel Raynal$^{\star,\ddag}$\\~\\
$^{\dag}$LINA, Universit\'e de Nantes, 44322 Nantes, France \\
$^{\star}$Institut Universitaire de France\\
$^{\ddag}$IRISA, Universit\'e de Rennes, 35042 Rennes, France \\
{\small {\tt  achour.mostefaoui@univ-nantes.fr~~ raynal@irisa.fr}}
~\\~\\ Tech Report \#2034, 15 pages, February 2016\\ 
IRISA, University of Rennes 1, France
}

\date{}

\maketitle


\begin{abstract}
  Atomic registers are certainly the most basic objects of computing
science.  Their implementation on top of an $n$-process asynchronous
message-passing system has received a lot of attention.  It has been
shown that $t<n/2$ (where $t$ is the maximal number of processes that
may crash) is a necessary and sufficient requirement to build an
atomic register on top of a crash-prone asynchronous message-passing system.
Considering such a context, this paper presents an algorithm which
implements a single-writer multi-reader atomic register with four message 
types only, and where no message needs to carry control information
in addition to its type. Hence, two bits are sufficient to capture all the
control information carried by all the implementation messages.  Moreover,
the messages of two types need to carry a data value while the messages 
of the two other types carry no value at all. As far as we know, this 
algorithm is the first with such an optimality property on the
size of control information carried by messages. It is also particularly 
efficient from a time complexity point of view. \\

\end{abstract}

~\\~\\~\\


\thispagestyle{empty}
\newpage
\setcounter{page}{1}

\section{Introduction}

Since Sumer time~\cite{K56}, and --much later-- Turing's machine
tape~\cite{T36}, read/write objects are certainly the most basic
communication objects.  Such an object, usually called a
{\it register}, provides its users (processes) with a write operation 
which defines the new value of the register, and a read operation which 
returns the value of the register. 
When considering sequential computing, registers are universal in the sense 
that they allow to solve any problem that can be solved~\cite{T36}.

\paragraph{Register in message-passing systems}
In a message-passing system, the computing entities communicate only
by sending and receiving messages transmitted through a communication network.
Hence, in such a system, a register is not a communication object
given for free, but constitutes a communication abstraction which must
be built with the help of the underlying communication network and 
the local memories of the processes.

Several types of registers can be defined according to which processes
are allowed to read or write the register, and the quality (semantics) of the
value returned by each read operation.  We consider here registers
which are single-writer multi-reader (SWMR), and atomic. 
Atomicity means that (a) each read or write operation appears
as if it had been  executed instantaneously at a single point of the time
line, between its start event and its end event, 
(b) no two operations appear at the same point of the time line, and 
(c) a read returns the value written
by the closest preceding write operation (or the initial value of the
register if there is no preceding write)~\cite{L86}.  Algorithms building
multi-writer multi-reader (MWMR) atomic registers from single-writer
single-reader (SWSR) registers with a weaker semantics (safe or regular
registers) have been introduced by L. Lamport in~\cite{L86,L86-2} 
(such  algorithms are described in several papers and 
textbooks, e.g.,~\cite{AW04,L96,R13-a,VA86}).

Many distributed algorithms have been proposed, which build a
register on top of a message-passing system, be it failure-free or
failure-prone. In the failure-prone case,  the addressed failure models 
are the process crash failure model, or the  Byzantine process failure 
model (see, the textbooks~\cite{AW04,L96,R10,R13}).  The most famous of 
these algorithms was proposed by H. Attiya, A. Bar-Noy,  and D. Dolev 
in~\cite{ABD95}.  This algorithm, which is usually called ABD
according to the names of its authors, considers an $n$-process asynchronous 
system in which up to $t<n/2$ processes may crash (it is also shown 
in~\cite{ABD95} that  $t<n/2$ is an upper bound 
of the number of process crashes which can be tolerated). 
This simple and elegant algorithm, 
relies on (a) quorums~\cite{V12}, and (b) a simple broadcast/reply 
communication pattern. ABD uses this pattern once in  a write operation, 
and twice in a read operation implementing an SWMR register
(informal presentations of ABD can be found in~\cite{A10,R08}).

\paragraph{Content of the paper}
ABD and  its successors (e.g.,~\cite{A00,MR16,V12}) associate an 
increasing  sequence number with each value that is written. 
This allows to easily identify each written value. 
Combined with the use of majority quorums, this value identification
allows each read invocation to return a value that satisfies the atomicity 
property (intuitively, a read always returns the ``last'' written value).  

Hence, from a communication point of view, 
in addition to the number of messages needed to implement a read
or a write operation, important issues are the number of different message 
types, and the size of the control information that each of them has to carry. 
As sequence numbers increase according to the number of write invocations, 
this number is not bounded, and the size of a message that carries a sequence
number can become arbitrarily large. 

A way to overcome this drawback consists in finding a modulo-based  
implementation of sequence numbers~\cite{IL93}, which can be used 
to implement read/write registers.  Considering this approach, one 
of the algorithms presented in~\cite{ABD95} uses messages that carry 
control information whose size is upper bounded by $O(n^5)$ bits 
(where $n$ is the total number of processes).  The algorithm presented 
in~\cite{A00} reduced this size to $O(n^3)$ bits. 
Hence the natural question: ``{\it How many bits of control information, 
a message has to carry, when one wants to implement an atomic read/write 
register?}''.

This is the question that gave rise to this paper, which shows that 
it is possible to implement an SWMR atomic register with four types of 
message carrying no control information in addition to their type. 
Hence, the result: {\it  messages carrying only two bits of control 
information are sufficient to implement an SWMR atomic register in the 
presence of asynchrony and up to $t<n/2$ unexpected process crashes}. 
Another important property of the proposed algorithm lies in its time 
complexity, namely, in a failure-free context and assuming a bound $\Delta$ 
on message transfer delays,  a write operation requires at most $2\Delta$ 
time units, and a read operation requires  at most $4\Delta$ time units.

\paragraph{Roadmap}
The paper is made up of~\ref{sec:conclusion} sections. The computing
model and the notion of an atomic register are presented in
Section~\ref{sec:model}.  The algorithm building an SWMR atomic
register, where messages carry only two bits of control information
(their type), in an asynchronous message-passing system prone to any
minority of process crashes is presented in Section~\ref{sec:algorithm}. 
Its proof appears in Section~\ref{sec:proof}. 
Finally, Section~\ref{sec:conclusion} concludes the paper.

\section{Computation Model and Atomic Read/Write Register}
\label{sec:model}

\subsection{Computation model}

\paragraph{Processes}
The computing model is composed of a set of $n$ sequential processes
denoted $p_1$, ..., $p_n$. Each process is asynchronous which means
that it proceeds at its own speed, which can be arbitrary and remains
always unknown to the other processes.  

A process may halt prematurely (crash failure), but executes correctly
its local algorithm until it possibly crashes. The model parameter $t$
denotes the maximal number of processes that may crash in a  run.
A process that crashes in a run is said to be {\it faulty}. Otherwise, 
it is {\it correct} or {\it non-faulty}.
Given a run,  ${\cal C}$ denotes the set of correct processes. 

\paragraph{Communication}
Each pair of processes  communicate  by sending and receiving messages through 
two uni-directional channels, one in each direction. Hence, the 
communication network is a complete network:
any process $p_i$  can directly send a message to any 
process $p_j$.
A process $p_i$ invokes 
the operation ``${\sf send}$ {\sc type}($m$) ${\sf to}$ $p_j$''
to send to $p_j$ the message $m$, whose type is  {\sc type}.
The operation ``${\sf receive}$ {\sc type}() ${\sf from}$ $p_j$'' 
allows $p_i$ to receive from $p_j$ a message  whose type is  {\sc type}. 

Each channel is reliable (no loss, corruption, nor creation of messages),
not necessarily first-in/first-out, and asynchronous (while the transit time 
of each message is finite, there is no upper bound on message transit times).

Let us notice that, due to process and message asynchrony, no process can
know if an other process crashed or is only very slow.

\paragraph{Notation} 
In the following, the previous computation model is denoted 
${\CAMP}_{n,t}[\emptyset]$ (unconstrained ${\cal C}$rash 
${\cal A}$synchronous ${\cal M}$essage-${\cal P}$assing).  

\subsection{Atomic read/write register}
\paragraph{Definition}
A {\it concurrent object} is an object that can be accessed by several 
processes (possibly simultaneously).  An SWMR {\it atomic} 
register (say $\REG$) is a concurrent object which provides exactly one 
process (called the writer) with an operation denoted $\REG.{\sf write}()$, 
and all processes with an operation denoted $\REG.{\sf read}()$.  
When the writer invokes $\REG.{\sf write}(v)$ it defines
$v$ as being the new value of $\REG$.  An SWMR atomic register  is
defined by the following set of properties~\cite{L86}.

\begin{itemize}
\vspace{-0.1cm}
\item Liveness. An invocation of an operation by a correct process terminates.
\vspace{-0.2cm}
\item Consistency (safety).  All the operations invoked by the
processes, except possibly --for each faulty process-- the last
operation it invoked, appear as if they have been executed
sequentially and this sequence of operations is such that:
\begin{itemize}
\vspace{-0.1cm}
\item each read returns the value written by the closest write that precedes
it (or the initial value of $\REG$ if there is no preceding write), 
\vspace{-0.1cm}
\item if an operation $op1$ terminates before an operation $op2$ starts, then 
 $op1$ appears before $op2$ in the sequence.
\end{itemize}
\end{itemize}

This set of properties states that, from an external observer point of
view, the read/write register appears as if it is accessed sequentially 
by the processes, and this sequence (a) respects the real time access order,
and (ii) belongs to the sequential specification of a register. 
More formal definitions can be found in~\cite{L86,M86}.
(When considering any object defined by a sequential
specification, atomicity is also called linearizability~\cite{HW90},
and it is then said that the object is {\it linearizable}.)

\paragraph{Necessary and sufficient condition} 
The constraint $(t<n/2)$ is a necessary and sufficient condition to implement 
an atomic read/write register in ${\CAMP}_{n,t}[\emptyset]$~\cite{ABD95}. 
Hence, the corresponding  constrained model is denoted ${\CAMP}_{n,t}[t<n/2]$.

\section{An Algorithm with Two-Bit Messages}
\label{sec:algorithm}

A distributed algorithm implementing an SWMR atomic register in 
${\CAMP}_{n,t}[t<n/2]$ is described in Figure~\ref{algo:two-bit-messages}. 
As already indicated, this algorithm uses only four types of messages, 
denoted {\sc write0}$()$, {\sc write1}$()$, 
{\sc read}$()$, and {\sc proceed}$()$. 
The messages {\sc write0}$()$ and {\sc write1}$()$ carry a data value, while 
the messages {\sc read}$()$ and {\sc proceed}$()$  carry only their type.  

\subsection{Notation and underlying principles}
\paragraph{Notation}
 $p_w$ denotes the writer process, $v_x$ denotes the $x^{\mathit th}$ value
written by $p_w$, and  $v_0$ is the initial value of the register $\REG$ 
that is built. 

\paragraph{Underlying principles}
The principle that underlies the algorithm is the following. 
First, each process (a) manages a local copy of the sequential history made up of
the values written by the writer, and (b) forwards, once to each process, 
each new value it learns. 
Then, in order that all processes obtain the same sequential history,
and be able to read up to date  values,  each process $p_i$ follows  
rules to forward a value to another process $p_j$, and manages accordingly
appropriate local variables, which store sequence numbers.  
\begin{itemize}
\vspace{-0.1cm}
\item 
Rule R1.
When, while it knows the first  $(x-1)$ written values, 
and only them, $p_i$ receives the $x^{\mathit th}$ written value, it forwards it 
to all the processes that, from its point of view, know  the first $(x-1)$ 
 written values  and no more. In this way, these processes will learn the 
$x^{\mathit th}$ written value (if not yet done when they receive the 
corresponding message forwarded by $p_i$). 
\vspace{-0.2cm}
\item
Rule R2.
The second forwarding rule is when $p_i$  receives the $x^{\mathit th}$ written 
value from a process $p_j$, while it knows the first $y$ written values, 
where $y>x$. 
In this case, $p_i$ sends the $(x+1)^{\mathit th}$ written value to $p_j$,
and only this value, in order $p_j$ increases its local sequential history with 
its next value (if not yet done when it receives the message from $p_i$). 
\vspace{-0.2cm}
\item
Rule R3.
To ensure a correct management of the local histories, 
and allow a process to help other processes in the construction of 
their  local histories (Rules R1 and R2), each process manages a 
sequence number-based local view of the progress of each other process
(as far as the  construction of their local history is concerned). 
\end{itemize}

As we are about to see, translating these rules into an algorithm,
provides us with a distributed algorithm where, while each process locally
manages sequence numbers, the only  control information carried by each 
message is its type, the number of different message  types being very 
small (namely $4$, as already indicated)\footnote{Such a constant number 
of message types is not possible from a ``modulo $f(n)$'' implementation of 
sequence numbers carried by messages. This is because, from a control 
information point of view,  each of the values in  
$\{0, 1,\ldots, f(n)-1\}$ defines a distinct message type.}.

\subsection{Local data structures}
Each process $p_i$ manages the following local data structures. 
\begin{itemize}
\vspace{-0.1cm}
\item $history_i$ is the prefix sequence of the values already written, as  
known by $p_i$;  $history_i$ is accessed with an array 
like-notation, and we have $history_i[0]=v_0$. 
As there is a single writer $p_w$,  
$history_w$ represents the history of the values written so far. 

\vspace{-0.2cm}
\item 
$w\_sync_i[1..n]$ is an array of sequence numbers;
$w\_sync_i[j]=\alpha$ means that, to $p_i$'s knowledge, $p_j$ knows
the prefix of $history_w$ until $history_w[\alpha]$. Hence,
$w\_sync_i[i]$ is the sequence number of the most recent value known by $p_i$,
and $w\_sync_w[w]$ is the sequence number of the last value written (by $p_w$).

\vspace{-0.2cm}
\item 
$r\_sync_i[1..n]$ is an array of sequence numbers; 
$r\_sync_i[j]=\alpha$ means that, to $p_i$'s knowledge, $p_j$ 
answered $\alpha$ of its read requests.
\vspace{-0.2cm}
\item $wsn$, $rsn$ and $sn$ are auxiliary local variables, the scope of 
each being restricted to the algorithm implementing an operation, 
or the processing  of a message, in which it occurs. 
\end{itemize}

\subsection{Channel behavior with respect to the message types 
{\sc write0}$()$ and {\sc write1}$()$}
\label{sec-alternating-bit}
As far as the messages {\sc write0}$()$ and {\sc write1}$()$ are concerned, 
the notation {\sc write}$(0,v)$ is used for {\sc write0}$(v)$, and similarly, 
{\sc write}$(1,v)$ is used for  {\sc write1}$(v)$.

When considering the two uni-directional channels connecting $p_i$ and $p_j$,
the algorithm, as we will see,
requires (a) $p_i$ to send to $p_j$ the sequence of messages 
{\sc write}$(1,v_1)$, {\sc write0}$(0,v_2)$, {\sc write}$(1,v_3)$, ...,  
{\sc write}$(x\mbox{ mod }2,v_x)$, etc., and (b)
$p_j$ to send to $p_i$ the very same sequence of messages 
{\sc write}$(1,v_1)$, {\sc write0}$(0,v_2)$, {\sc write}$(1,v_3)$, ...,  
{\sc write}$(x\mbox{ mod }2,v_x)$, etc.

Moreover, the algorithm forces process $p_i$ to send to $p_j$ the message 
{\sc write}$(x\mbox{ mod }2,v_x)$, only when it has received from $p_j$ the 
message {\sc write}$((x-1)\mbox{ mod }2,v_{x-1})$. 
From the point of view of the write messages, these communication rules 
actually implement the {\it alternating bit}  protocol~\cite{BSW69,L68}, 
which ensures the following properties: 
\begin{itemize}
\vspace{-0.2cm}
\item Property P1: each of the two uni-directional channels connecting $p_i$ 
and $p_j$ allows at most one message {\sc write}$(-,-)$ to bypass another
message {\sc write}$(-,-)$, which, thanks to the single control bit carried by 
these messages allows the destination process (e.g., $p_i$) 
to process the messages  {\sc write}$(-,-)$ it receives from (e.g., $p_j$) 
in their sending order. 
\vspace{-0.2cm}
\item  Property P2: $p_i$ and $p_j$ are synchronized in such a way that 
$0\leq |w\_sync_i[j] - w\_sync_j[i]| \leq 1$. This is the translation
of Property P1 in terms of the pair of local synchronization-related variables 
$\langle w\_sync_i[j], w\_sync_j[i]\rangle$.  
\end{itemize}
Let us insist on the fact that this ``alternating bit'' message exchange
pattern is only on the write messages. It imposes no constraint on the 
messages of the types {\sc read}$()$ and {\sc proceed}$()$ exchanged 
between $p_i$ and $p_j$, which can come in between, at any place in the 
sequence of the write messages sent by a process $p_i$ to a process $p_j$. 

\subsection{The algorithm implementing the ${\sf write}()$ operation}

This algorithm is described at lines~\ref{NC-SWMR-01}-\ref{NC-SWMR-04},
executed by the writer $p_w$, and line~\ref{NC-SWMR-11}-\ref{NC-SWMR-18},
executed by any process. 

\paragraph{Invocation of the operation ${\sf write}()$}
When $p_w$ invokes ${\sf write}(v_x)$ (we have then $w\_sync_w[w]=x-1$),
it increases $w\_sync_w[w]$ and writes $v_x$ at the tail of its
local history variable (line~\ref{NC-SWMR-01}). This value is locally 
identified by its sequence number $x=wsn$. 

Then $p_w$ sends the message {\sc write}$(b,v_x)$,
 where $b= (wsn \mbox{ mod } 2)$, to each  process $p_j$ that (from its point 
of view) knows all the previous write invocations, and only to these 
processes.  
According to the definition of $w\_sync_w[1..n]$, those are the processes 
$p_j$ such that $w\_sync_w[j]=wsn-1= w\_sync_w[w]-1$ (line~\ref{NC-SWMR-02}).
Let us notice that this ensures the requirement $p_i$ needs to satisfy 
when it sends a message in order to benefit from 
the properties provided by the alternating bit communication pattern.

Finally, $p_w$ waits until it knows that a quorum of at least $(n-t)$
processes knows the value $v_x$ is it writing. The fact that a process
$p_j$ knows this $x^{\mathit th}$  value is captured by the predicate 
$w\_sync_w[j]=wsn(=x)$ (line~\ref{NC-SWMR-03}).

\begin{figure*}[th]
\centering{\fbox{
\begin{minipage}[t]{150mm}
\footnotesize
\renewcommand{\baselinestretch}{2.5}
\resetline
\begin{tabbing}
aaaaa\=a\=aaaaa\=aaaaaa\=\kill

{\bf local variables initialization:}\\
\> $history_i[0]\gets v_0$;
   $w\_sync_i[1..n] \gets [0,\ldots,0]$;  
   $r\_sync_i[1..n] \gets [0,\ldots,0]$.\\~\\

{\bf operation} $\mathsf{write}$($v$) {\bf is} 
\% invoked by $p_i=p_w$ (the writer) \% \\

\line{NC-SWMR-01} 

\> $wsn \leftarrow w\_sync_w[w] +1$; 
   $w\_sync_w[w]\leftarrow wsn$; 
   $history_w[wsn]\leftarrow v$; $b \leftarrow wsn \mbox{ mod } 2$;  \\

\line{NC-SWMR-02}
\> {\bf for each} $j$ such that $w\_sync_w[j] = wsn-1$ 
    {\bf do} $\mathsf{send}$ {\sc write}($b,v$) ${\sf to}$ $p_j$ 
   {\bf end for};\\

\line{NC-SWMR-03} 
\> ${\sf wait}$ 
 \big($z  \geq (n-t)$ where $z$ is the number of processes $p_j$ 
                      such that $w\_sync_w[j] = wsn$\big);\\
  
\line{NC-SWMR-04} 
\>  ${\sf return}()$\\
{\bf end operation}.\\~\\

{\bf operation} $\mathsf{read}$() {\bf is}
\% the writer can directly returns $history_i[w\_synch_i[i]]$ \% \\

\line{NC-SWMR-05} 
\> $rsn\leftarrow r\_sync_i[i]+1$;  $r\_sync_i[i]\leftarrow rsn$; \\

\line{NC-SWMR-06} 
\> {\bf for each} $j\in\{1,...n\} \setminus \{i\}$
   {\bf do} $\mathsf{send}$ {\sc read}$()$ ${\sf to}$ $p_j$ {\bf end for};\\

\line{NC-SWMR-07} 
\> ${\sf wait}$ 
 \big($z \geq (n-t)$ where $z$ is the number of processes $p_j$ such that
                   $r\_sync_i[j] = rsn$\big); \\

\line{NC-SWMR-08} 
\> let $sn = w\_sync_i[i]$; \\

\line{NC-SWMR-09} 
\> ${\sf wait}$ 
 \big($z\geq (n-t)$ where $z$ is the number of processes $p_j$ 
                  such that $w\_sync_i[j] \geq sn$\big); \\

\line{NC-SWMR-10} 
\>  ${\sf return}(history_i[sn])$\\
{\bf end operation}. \\
\%-------------------------------------------------------------------------------------------------------------------------------------\\~\\

{\bf when} {\sc write}$(b,v)$ {\bf is received from} $p_j$ {\bf do}\\

\line{NC-SWMR-11} 
\>  ${\sf wait}$  $\big(b=  (w\_sync_i[j] +1)\mbox { mod }2 \big)$;\\

\line{NC-SWMR-12} 
\>  $wsn \leftarrow w\_sync_i[j] +1$; \\

\line{NC-SWMR-13} 
\> {\bf if}  \= ($wsn = w\_sync_i[i]+1$) \\

\line{NC-SWMR-14} 
  \> \>   {\bf then} \= $w\_sync_i[i]\leftarrow wsn$; 
                        $history_i[wsn]\leftarrow v$;
                        $b \leftarrow wsn \mbox{ mod } 2$;  \\

\line{NC-SWMR-15}
\>\> \>  {\bf for each} $\ell$ such that $w\_sync_i[\ell] = wsn -1$ 
         {\bf do} $\mathsf{send}$ {\sc write}($b,v$)  ${\sf to}$ $p_\ell$ 
         {\bf end for}\\

\line{NC-SWMR-16}
\>\>    {\bf else} \> {\bf if}  \= ($wsn < w\_sync_i[i]$) 
        {\bf then} $b \leftarrow (wsn+1)  \mbox{ mod } 2$; 
        $\mathsf{send}$ {\sc write}($b,history_i[wsn+1]$) ${\sf to}$ $p_j$
        {\bf end if} \\

\line{NC-SWMR-17} \> {\bf end if};\\ 

\line{NC-SWMR-18} $w\_sync_i[j]\leftarrow wsn$. \\~\\

{\bf when}
{\sc read}$()$ {\bf is received from} $p_j$ {\bf do}\\

\line{NC-SWMR-19} 
\>  $sn \leftarrow  w\_sync_i[i]$; \\

\line{NC-SWMR-20} 
\> ${\sf wait}$ $(w\_sync_i[j]\geq sn)$; \\

\line{NC-SWMR-21} 
\> $\mathsf{send}$ {\sc proceed}() $\mathsf{to}$ $p_j$. \\~\\

{\bf when} 
{\sc proceed}() {\bf is received from} $p_j$ {\bf do}\\

\line{NC-SWMR-22} 
\> $r\_sync_i[j] \gets r\_sync_i[j]+1$.

\end{tabbing}
\end{minipage}
}
\caption{Single-writer multi-reader atomic register in 
        ${\cal CAMP}_{n,t}[t<n/2]$ with counter-free messages}
\label{algo:two-bit-messages}
}
\end{figure*}
 
\paragraph{Reception of a message {\sc write}$(b,v)$ from a process $p_j$}
When $p_i$ receives a message {\sc write}$(b,v)$ from a process $p_j$,
it first waits until the waiting
predicate of line~\ref{NC-SWMR-11} is satisfied. This waiting statement is 
nothing else than the the reception part of the alternating bit 
algorithm, which guarantees that the messages  {\sc write}$()$  from $p_j$ 
are processed in their sending order. 
When, this waiting predicate is satisfied, all messages sent by $p_j$  before 
{\sc write}$(b,v)$ have been received and  processed by $p_i$, and 
consequently the message {\sc write}$(b,v)$  is the $swn^{\mathit th}$ message 
sent by $p_j$ (FIFO order), where $wsn=w\_sync_i[j]+1$, which means that 
$history_j[wsn]=v$ (line~\ref{NC-SWMR-12}).  

When this occurs,  $p_i$ learns that $v$ is the next value to be added to its 
local history if additionally we have $w\_sync_i[i]=wsn-1$. In this case
(predicate of line~\ref{NC-SWMR-13}), $p_i$ (a) adds $v$ at the tail
of its history (line~\ref{NC-SWMR-14}), and (b) forwards the message
{\sc write}$(b,v)$ to the processes that, from its local point of view,
know the first $(wsn-1)$ written values and no more
(line~\ref{NC-SWMR-15}, forwarding Rule R1).

If $wsn < w\_sync_i[i]$, from $p_i$'s local point of view,  the history known
by $p_j$ is a strict prefix of its own history. Consequently, $p_i$ sends
to $p_j$ the message {\sc write}$(b',v')$, where 
$b'=((wsn+1)\mbox{ mod } 2)$ and  $v'=history_i[wsn+1]$
(line~\ref{NC-SWMR-16} applies the forwarding Rule R2 in order
to allow  $p_j$ to catch up its lag, if not yet done when it will receive 
the message {\sc write}$(b',v')$ sent by $p_i$).  Finally,
as $p_j$ sends to $p_i$ a single message per write operation, whatever 
the value of $wsn$, $p_i$ updates $w\_sync_i[j]$ (line~\ref{NC-SWMR-18}).
 
\paragraph{Remark}
As far as the written values are concerned, the algorithm implementing 
the operation ${\sf write}()$ can be seen as a fault-tolerant ``synchronizer'' 
(in the spirit of~\cite{A85}), which ensures the mutual consistency of  
the local histories between any two neighbors with the help  of an 
alternating bit algorithm executed by each pair of neighbors~\cite{BSW69,L68}.  

\subsection{The algorithm implementing the ${\sf read}()$ operation}
This algorithm is described at lines~\ref{NC-SWMR-05}-\ref{NC-SWMR-10}
executed by a reader $p_i$, and lines~\ref{NC-SWMR-19}-\ref{NC-SWMR-22}
executed by any  process. 

\paragraph{Invocation of the operation ${\sf read}()$}
The invoking process $p_i$ first increments its local read request
sequence number $r\_sync_i[i]$ and broadcasts its read request in a
message {\sc read}$()$, which carries neither additional control information, 
nor a data value  (lines~\ref{NC-SWMR-05}-\ref{NC-SWMR-06}).  If $p_i$ crashes
during this broadcast, the message {\sc read}$()$ is received by an
arbitrary subset of processes (possibly empty). Otherwise, $p_i$ waits
until it knows that at least $(n-t)$ processes received its
current request (line~\ref{NC-SWMR-07}).

When this occurs, $p_i$ considers the sequence number of the 
last value in its history, namely $sn=w\_sync_i[i]$ (line~\ref{NC-SWMR-08}).
This is the value it will return, namely  $history_i[sn]$  
(line~\ref{NC-SWMR-10}). But in order to ensure atomicity, 
before returning  $history_i[sn]$, $p_i$ waits
until at least $(n-t)$ processes know this value (and may be more).
From $p_i$'s point of view, the corresponding waiting predicate  translates 
in ``at least  $(n-t)$ processes $p_j$ are such that $w\_sync_i[j]\geq sn$''. 
 
\paragraph{Reception of a message {\sc read}$()$ sent by  a process $p_j$}
When a process $p_i$ receives a message {\sc read}$()$ from a process $p_j$
(hence, $p_j$ issued a read operation),  
it considers the most recent written value it  knows (the sequence number
of this value is $sn=w\_sync_i[i]$, line~\ref{NC-SWMR-19}), and waits until 
it knows that $p_j$ knows this value, which is locally captured by the 
sequence number-based predicate $w\_sync_i[j]\geq sn$ (line~\ref{NC-SWMR-20}).
When this occurs, $p_i$ sends the message {\sc proceed}$()$ to $p_j$ 
which is  allowed  to progress as far as $p_i$ is concerned. 

The control messages {\sc read}$()$ and {\sc proceed}$()$ (whose
sending is controlled by a predicate) implement a synchronization which
--as far as $p_i$ is concerned-- forces the reader process $p_j$ to wait until 
it knows a ``fresh'' enough value, where ``freshness'' is locally defined 
by $p_i$ as the  last value it was knowing when it received the message
{\sc read}$()$ from $p_j$  (predicate of line~\ref{NC-SWMR-20}).

\paragraph{Reception of a message {\sc proceed}$()$  sent by a process $p_j$}
When $p_i$ receives a message {\sc proceed}$()$ from a process $p_j$, it learns 
that its local history is as fresh as $p_j$'s history when $p_j$ received 
its message {\sc read}$()$. Locally, this is captured  by the incrementation 
of $r\_sync_i[j]$, namely $p_j$ answered all the read requests of $p_i$ until 
the  $(r\_sync_i[j])^{\mathit th}$ one.

\section{Proof of the Algorithm}
\label{sec:proof}
Let us remind that $\cal C$ is the set of correct processes, 
$p_w$ the writer, and $v_x$ the $x^{\mathit th}$ value written by $p_w$. 

\begin{lemma}
\label{lemma-wsync-incrementation}
$\forall i,j$: $w\_sync_i[j]$  increases by steps  equal to $1$.
\end{lemma}
As this lemma is used in all other lemmas,  it will not be explicitly 
referenced. 

\begin{proofL}
Let us first observe that, due to the sending predicates 
of line~\ref{NC-SWMR-02} (for the writer), and lines~\ref{NC-SWMR-15} 
and~\ref{NC-SWMR-16} for any process $p_i$, no process sends a message 
{\sc write}$(-,-)$ to itself. 
  
As far as $w\_sync_i[i]$ is concerned, and according to the previous 
observation,we have the following.
The writer increases $w\_sync_w[w]$ only at line~\ref{NC-SWMR-01}. 
Any reader process $p_i$  increases  $w\_sync_i[i]$  at 
line~\ref{NC-SWMR-14}, and due to  line~\ref{NC-SWMR-12}
and  the predicate of line~\ref{NC-SWMR-13}, the increment is $1$. 
Let us now consider the case  of $w\_sync_i[j]$ when $i\neq j$. 
An incrementation of such a local variable occurs only at 
line~\ref{NC-SWMR-18}, where (due to line~\ref{NC-SWMR-12}) we have
$wsn=w\_sync_i[j]+1$, and the lemma follows.  
\renewcommand{\toto}{lemma-wsync-incrementation}
\end{proofL}

\begin{lemma}
\label{lemma-wsync-i-wrt-j}
$\forall i,j:~ w\_sync_i[i] \geq w\_sync_j[i]$. 
\end{lemma}

\begin{proofL}
Let us first observe, that the predicate is initially true.  Then, a
local variable $w\_sync_j[i]$ is increased by $1$, when $p_j$
receives a message {\sc write}$(-,-)$ from $p_i$ (lines~\ref{NC-SWMR-12}
and~\ref{NC-SWMR-18}).  Process $p_i$ sent this message at
line~\ref{NC-SWMR-02}  or~\ref{NC-SWMR-16} if $i=w$, and  at 
lines~\ref{NC-SWMR-15} or~\ref{NC-SWMR-16} for any $i\neq w$. 
If the sending of the message  {\sc write}$(b,-)$ by $p_i$ occurs at
line~\ref{NC-SWMR-02} or~\ref{NC-SWMR-15}, $p_i$ increased $w\_sync_i[i]$ 
at the previous line. If the sending occurs at line~\ref{NC-SWMR-16}, 
$w\_sync_i[i]$ was increased during a previous message reception.
\renewcommand{\toto}{lemma-wsync-i-wrt-j}
\end{proofL}

\begin{lemma}
\label{lemma-wsync-max}
$\forall i$: $w\_sync_i[i]=\max \{w\_sync_i[j]\}_{1\leq j\leq n}$.
\end{lemma}

\begin{proofL}
The lemma is trivially true for the writer process $p_w$.
Let us consider any other process $p_i$, different from $p_w$. The proof is 
by induction on the number of messages  {\sc write}$(-,-)$ received by $p_i$.
Let  $P(i,m)$ be the predicate 
$w\_sync_i[i]=\max \{w\_sync_i[j]\}_{1\leq j\leq n}$, 
where $m$ is the number of messages  {\sc write}$(-,-)$ processed by $p_i$.
The predicate $P(i,0)$ is  true.
Let us assume $P(i,m')$ is true for any $m'$ such that $0\leq m'\leq m$. 
Let $p_j$ be the process that sends to $p_i$ the $(m+1)^{th}$ message 
{\sc write}$(b,-)$, and let  $w\_sync_i[i]=x$ when $p_i$ starts processing 
this message.  There are four  cases to consider. 
\begin{itemize}
\vspace{-0.1cm}
\item Case 1. 
When the message {\sc write}$(-,-)$  from $p_j$ is processed by $p_i$, 
we have  $w\_sync_i[i]+1=w\_sync_i[j]+1$. 
As  the predicate of line~\ref{NC-SWMR-13} is satisfied when this message 
is processed, $p_i$ updates  $w\_sync_i[i]$ to the value $(x+1)$ at 
line~\ref{NC-SWMR-14}. Moreover, it also  updates $w\_sync_i[j]$ to the 
same value $(x+1)$ at line~\ref{NC-SWMR-18}.  
As $P(i,m)$  is true, it follows that $P(i,m+1)$ is true after $p_i$ 
processed  the message.

\vspace{-0.1cm}
\item Case 2. 
When the message {\sc write}$(-,-)$  from $p_j$  is processed by  $p_i$, 
we have  $w\_sync_i[j]+1< w\_sync_i[i]=x$. In this case,  $p_i$ does not modify
$w\_sync_i[i]$. It only updates $w\_sync_i[j]$  to its next value 
(line~\ref{NC-SWMR-18}), which is smaller than $x$. As $P(i,m)$ is true, 
it follows that $P(i,m+1)$ is true after $p_i$ processed  the message.

\vspace{-0.1cm}
\item Case 3. 
When the message {\sc write}$(-,-)$ from $p_j$ is processed by $p_i$, 
we have  $w\_sync_i[j]+1 = w\_sync_i[i]=x$. In this case,
both the predicates of lines~\ref{NC-SWMR-13} and~\ref{NC-SWMR-16}  
are false. It follows that $p_i$ executes only the update of 
line~\ref{NC-SWMR-18},  and we have then  $w\_sync_i[j] = w\_sync_i[i]=x$.
As $P(i,m)$ is true, $P(i,m+1)$ is true after $p_i$ processed  the message.

\vspace{-0.1cm}
\item Case 4.  
When the message {\sc write}$(-,-)$ from $p_j$ is processed by  $p_i$, 
we have  $w\_sync_i[j]+1 > w\_sync_i[i]+1 = x+1$. 
In this case, due to (a) $w\_sync_i[j] \leq  w\_sync_i[i]$ 
(induction assumption satisfied when the 
message {\sc write}$(-,-)$ arrives at $p_i$  from $p_j$), and 
(b) the fact that $w\_sync_i[j]$ increases by step $1$ 
(Lemma~\ref{lemma-wsync-incrementation}), we necessarily have 
$w\_sync_i[i] +1 \geq w\_sync_i[j]+1$, when the message is received. 
Hence, we obtain  $w\_sync_i[j]+1 > w\_sync_i[i] +1 \geq w\_sync_i[j]+1$,
a contradiction. It follows that this case cannot occur.
\end{itemize}
\vspace{-0.5cm}
\renewcommand{\toto}{lemma-wsync-max}
\end{proofL}

\begin{lemma}
\label{lemma-prefix-history}
$\forall i$: $history[0..w\_sync_i[i]]$ is a prefix of 
$history[0..w\_sync_w[w]]$.
\end{lemma}

\begin{proofL}
The proof of this lemma rests on the properties P1 and P2 provided 
by the underlying ``alternating bit'' communication pattern imposed 
on the messages {\sc write}$(-,-)$ exchanged by any pair of processes 
$p_i$ and $p_j$.  If follows from these properties  (obtained from the use 
of parity bits carried by every message {\sc write}$(-,-)$,  and the 
associated wait statement of line~\ref{NC-SWMR-11}) that, 
$p_i$ sends to $p_j$ the message {\sc write}$(-,v_x)$,  only after 
it knows that $p_j$ received {\sc write}$(-,v_{x-1})$. 
Moreover, it follows from the management of the local sequence numbers
$w\_sync_i[1..n]$, that no process sends twice the same message 
{\sc write}$(-,v_x)$. Finally, due to the predicate of line~\ref{NC-SWMR-11},
two consecutive  messages {\sc write}$(0,-)$ and {\sc write}$(1,-)$ sent by a 
process $p_i$ to a process $p_j$  are processed in their sending order.

The lemma then follows from these properties, and the fact that, 
when at lines~\ref{NC-SWMR-13}-\ref{NC-SWMR-14} 
a process $p_i$  assigns a value $v$  to $history_i[x]$, this value 
was carried by  $x^{\mathit th}$ message {\sc write}$(-,v)$ 
sent by some process $p_j$,  and is the value of  $history_j[x]$.
It follows that no two processes have different histories, 
from which we conclude that   $history_i[x]= history_w[x]$. 
\renewcommand{\toto}{lemma-prefix-history}
\end{proofL}

\begin{lemma}
\label{lemma-eventual-history}
$\forall i \in {\cal C}, \forall j:$ we have:\\
{\em R1:}  $(w\_sync_i[i]=w\_sync_i[j]=x) \Rightarrow p_i$ sent $x$  messages
{\sc write}$(-,-)$  to $p_j$,\\
{\em R2:} $(w\_sync_i[i]>w\_sync_i[j]=x) \Rightarrow p_i$ sent $x+1$  messages
{\sc write}$(-,-)$ to $p_j$.
\end{lemma}

\begin{proofL}
Both predicates are  initially true ($w\_sync_i[i]=w\_sync_i[j]=0$ and 
no message was previously sent by $p_i$ to $p_j$). 
The variables involved in the premises of the predicates R1 and R2 can be 
modified in the execution of a write operation (if $p_i$ is the writer), or 
when a message {\sc write}$(-,-)$ arrives at process $p_i$ from process $p_j$. 
Let us suppose that R1 and R2 are true until the value $x$, and let us 
show that they remain true for the value $(x+1)$. 

During the execution of a write operation, if
$w\_sync_w[w]=w\_sync_w[j]=x$, the local variable $w\_sync_w[w]$ is 
incremented to $(x+1)$, and the $(x+1)^{th}$ message {\sc write}$(-,-)$ is sent 
by $p_w$ to $p_j$  (lines~\ref{NC-SWMR-01}-\ref{NC-SWMR-02}). 
R1 and R2 remain true. 
If $w\_sync_w[w]>w\_sync_w[j]=x$,  the local variable $w\_sync_w[w]$ is
incremented  at line~\ref{NC-SWMR-01}, but  no message is sent to $p_j$
at line~\ref{NC-SWMR-02}, which falsifies neither R1 nor R2.\\

When a process $p_i$ receives a message {\sc write}$(-,-)$  from 
a process $p_j$, there are also two cases, according to the values of  
$w\_sync_i[i]$ and $w\_sync_i[j]$ when $p_i$ starts processing the message
at line~\ref{NC-SWMR-12}. 
\begin{itemize}
\vspace{-0.2cm}
\item 
Case 1.  $w\_sync_i[i]=w\_sync_i[j]=x$. 
In this case,  the predicate of line~\ref{NC-SWMR-13} is satisfied. 
It follows that both   $w\_sync_i[i]$ and  $w\_sync_i[j]$ are  incremented 
to $(x+1)$ (at line~\ref{NC-SWMR-14} for  $w\_sync_i[i]$ and
line~\ref{NC-SWMR-18} for $w\_sync_i[j]$). 
Moreover, when $p_i$ executes line~\ref{NC-SWMR-15} we have  
$w\_sync_i[i]=w\_sync_i[j]-1$, and consequently $p_i$ sends 
a  message  {\sc write}$(-,-)$ to $p_j$ (the fact this 
message is the $(x+1)^{\mathit th}$ follows from the induction assumption). 
Hence, R1 and R2 are true
when $p_i$ terminates the processing of the message {\sc write}$(-,-)$
received from $p_j$. 
\vspace{-0.2cm}
\item Case $w\_sync_i[i]>w\_sync_i[j]=x$. In this case, $w\_sync_i[j]$ is 
incremented to $x+1$ at line~\ref{NC-SWMR-18}, while   $w\_sync_i[i]$ 
is not (because the predicate of line~\ref{NC-SWMR-13} is false). 
Two sub-cases are considered
according to the values of  $w\_sync_i[i]$ and $w\_sync_i[j]$.
\begin{itemize}
\vspace{-0.2cm}
\item 
If $w\_sync_i[i]=x+1$ (this is the value $w\_sync_i[j]$ will obtain at
line~\ref{NC-SWMR-18}), the predicate of line~\ref{NC-SWMR-16} 
is  false, and  no message is sent to $p_j$. 
R1 and R2 remains true, as, by the induction assumption, 
$p_i$ already sent $(x+1)$  messages {\sc write}$(-,-)$.
\vspace{-0.1cm}
\item 
If $w\_sync_i[i]>x+1$, the predicate of line~\ref{NC-SWMR-16} 
is satisfied, and the $(x+2)^{th}$ message {\sc write}$(-,-)$ is sent to 
$p_j$ at this line, maintaining satisfied the predicates R1 and R2.
\end{itemize}
\end{itemize}
\vspace{-0.5cm}
\renewcommand{\toto}{lemma-eventual-history}
\end{proofL}

\begin{lemma}
\label{lemma-i-j-sn}
$\forall i,j\in {\cal C}$, if $w\_sync_i[i] = x$, there is a finite  time
after which $w\_sync_i[j]\geq x$.
\end{lemma}

\begin{proofL}
Let us first notice that, due to Lemma~\ref{lemma-synchronizer},
all {\sc write}$(-,-)$ messages received by correct processes will
eventually satisfy the predicate line~\ref{NC-SWMR-11} and will be processed. 

The proof is by contradiction. Let us assume that there exists some correct 
process $p_j$ such that $w\_sync_i[j]$ stops increasing forever at some 
value $y<x$. Let us first notice
that there is no message {\sc write}$(-,-)$ in transit from $p_j$ to 
$p_i$ otherwise its reception by $p_i$ will entail the incrementation of
$w\_sync_i[j]$ from $y$ to $y+1$, contradicting the assumption. 
So, let us consider the last message {\sc write}$(-,-)$ sent by $p_j$ 
to $p_i$ and processed by $p_i$. There are three cases to consider when this 
message is received by $p_i$  at line \ref{NC-SWMR-11}.
(Let us remind that, due to to Lemma \ref{lemma-wsync-max}, 
$w\_sync_i[i] \geq w\_sync_i[j]$.) 
\begin{itemize}
\vspace{-0.2cm}
\item 
Case 1. $w\_sync_i[i] = w\_sync_i[j]=y-1<x-1$. The variables $w\_sync_i[i]$ 
and $w\_sync_i[j]$ are both incremented at lines~\ref{NC-SWMR-14} 
and~\ref{NC-SWMR-18} respectively to the value $y<x$. 
As by assumption, $w\_sync_i[i]$ will attain the value $x$, 
it will be necessarily  incremented in the future to reach $x$. 
The next time $w\_sync_i[i]$ is incremented, a message {\sc write}$(-,-)$ 
is sent by $p_i$ to $p_j$  (at line~\ref{NC-SWMR-15}). Due to
Lemma \ref{lemma-eventual-history}, $p_i$ sent $y+1$ messages  
{\sc write}$(-,-)$ to $p_j$ and eventually $w\_sync_j[i]$ will be equal 
to $y+1$. When the last of these messages arrives and is processed by $p_j$, 
there are two cases. 
\begin{itemize}
\vspace{-0.2cm}
\item 
Case  $w\_sync_j[j]=y$ (as $p_i$ sent $y+1$ messages  {\sc write}$(-,-)$ 
to $p_j$, $w\_sync_j[j]$ cannot be smaller than $y$).
In this case, $w\_sync_j[j]=y$   is increased,
and a message {\sc write}$(-,-)$ is necessarily sent by $p_j$ to $p_i$ 
(line~\ref{NC-SWMR-15}). This contradicts the assumption that the message 
we considered was the last message sent by $p_j$ to $p_i$.
\vspace{-0.1cm}
\item 
Case  $w\_sync_j[j]\geq y+1$. In this case, 
as $p_i$ sent previously $y$ messages to $p_j$, we 
necessarily have  $w\_sync_j[i]=y$. In this case, the predicate of 
line~\ref{NC-SWMR-13} is false, while the one of 
line~\ref{NC-SWMR-16} is satisfied. Hence, $p_j$ sends 
a message {\sc write}$(-,-)$ to $p_i$. A contradiction.
\end{itemize}
\vspace{-0.2cm}
\item 
Case 2. $w\_sync_i[i] = w\_sync_i[j]+1=y<x$. 
In this case, when $p_i$ receives the last message 
{\sc write}$(-,-)$ from $p_j$,  the variable $w\_sync_i[j]$ is 
incremented at line~\ref{NC-SWMR-18} to  the value $y<x$. Moreover, 
by the contradiction assumption, no more message {\sc write}$(-,-)$ is 
sent by $p_j$ to $p_i$. 

Hence, we have now
$w\_sync_i[i] = w\_sync_i[j]=y<x$, and the variable $w\_sync_i[i]$ will be
incremented in the future to reach $x$. A reasoning similar to the previous 
one shows that $p_j$ will send a message {\sc write}$(-,-)$ to $p_i$ in the 
future, which contradicts the initial assumption. 
\vspace{-0.2cm}
\item 
Case 3. $w\_sync_i[i] > w\_sync_i[j]+1$. The reception by $p_i$ of the  
last message  {\sc write}$(-,-)$ from $p_j$  entails the incrementation of  
$w\_sync_i[j]$ to its next value. However as
$w\_sync_i[i] > w\_sync_i[j]$ remains true, a message {\sc write}$(-,-)$ 
is sent by $p_i$  to $p_j$ at line~\ref{NC-SWMR-16}. 
Similarly to the previous cases, the reception of this message by $p_j$ 
will direct it to send another message  {\sc write}$(-,-)$ to $p_i$,
contradicting the initial assumption.
\end {itemize}
Hence, $w\_sync_i[j]$ cannot stop increasing before
reaching $x$, which proves the lemma.
\renewcommand{\toto}{lemma-i-j-sn}
\end{proofL}

\begin{lemma}
\label{lemma-synchronizer}
No correct process blocks forever at line~{\em\ref{NC-SWMR-11}}.
\end{lemma}

\begin{proofL}
The fact that the waiting  predicate of line~\ref{NC-SWMR-11}
is eventually satisfied follows from the following observations.
\begin{itemize}
\vspace{-0.2cm}
\item  As the network is reliable, all the messages that are 
sent are received. Due to lines~\ref{NC-SWMR-02}
and~\ref{NC-SWMR-15}-\ref{NC-SWMR-16}, this means that, for any $x$,  if 
{\sc write}$(-,v_x)$ is received while $m$ = {\sc write}$(-,v_{x-1})$  has not, 
then  $m$ will be eventually received. 
\vspace{-0.2cm}
\item  The message exchange pattern involving  any two
messages {\sc write}$(0,-)$ and  {\sc write}$(1,-)$ (sent consecutively) 
exchanged between each pair  of processes is the ``alternating bit pattern'', 
from which it follows that  no two messages {\sc write}$(b,-)$ 
(with the same $b$) can be received consecutively. 
\vspace{-0.2cm}
\item
It follows that the predicate of line~\ref{NC-SWMR-11}
is a simple re-ordering predicate for any pair of  messages 
such that {\sc write}$(-,v_{x})$ was received before 
{\sc write}$(-,v_{x-1})$. When this predicate  
is not satisfied for a message $m$ = {\sc write}$(b,-)$, 
this is because a message $m'$ = {\sc write}$(1-b,-)$, 
will necessarily arrive and  be  processed before $m$. 
After that, the predicate of line~\ref{NC-SWMR-11} becomes true for $m$. 
\end{itemize}
\vspace{-0.4cm}
\renewcommand{\toto}{lemma-synchronizer}
\end{proofL}

\begin{lemma}
\label{lemma-write-termination}
If the writer does not crash during a write operation, it terminates it. 
\end{lemma}

\begin{proofL}
Let us first notice that, due to Lemma~\ref{lemma-synchronizer}, 
the writer cannot block forever at line~\ref{NC-SWMR-11}. 

When it invokes a new write operation, the writer $p_w$ first increases 
the write sequence number $w\_sync_w[w]$ to its next value $wsn$ 
(line~\ref{NC-SWMR-01}).  
If $p_w$  does not crash, it follows from Lemma~\ref{lemma-i-j-sn} 
that we eventually have $w\_sync_i[i]\geq w\_sync_w[i] = wsn$ at each 
correct process $p_i$. Consequently, the writer cannot block forever
at line~\ref{NC-SWMR-03} and the lemma follows.
\renewcommand{\toto}{lemma-write-termination}
\end{proofL}

\begin{lemma}
\label{lemma-read-termination}
If a process does not crash during a read operation, it terminates it. 
\end{lemma}

\begin{proofL}
Let us first notice that, due to Lemma~\ref{lemma-synchronizer}, 
the reader cannot block forever at line~\ref{NC-SWMR-11}. 

Each time a process $p_i$ executes a read operation it broadcasts a
message {\sc read}$()$ to all the other processes  (line \ref{NC-SWMR-06}).
Let us remind that its local variable  $r\_sync _i[i]$ counts the
number of messages {\sc read}$()$ it has broadcast,
while  $r\_sync _i[j]$ counts the number of  messages {\sc proceed}$()$ 
it has received from $p_j$ (line \ref{NC-SWMR-22}) in response to its
{\sc read} messages {\sc read}$()$.

When the predicate of line~\ref{NC-SWMR-07} becomes true at the reader 
$p_i$, there are at least $(n-t)$ processes that answered  the 
$r\_sync_i[i]$ messages {\sc read}$()$ it sent 
(note that $r\_sync_i[i]$ is incremented line \ref{NC-SWMR-05}
and $p_i$ does not send messages {\sc read}$()$ to itself).
We claim that each message {\sc read}$()$ sent by $p_i$ to a 
correct process $p_j$ is eventually acknowledged by a 
a message {\sc proceed}$()$ send by $p_j$ to $p_i$. 
It follows from this claim and line~\ref{NC-SWMR-22} 
executed by $p_i$ when it receives a message {\sc proceed}$()$, that 
the predicate of line~\ref{NC-SWMR-07} is eventually satisfied, 
and consequently, $p_i$ cannot block forever at line~\ref{NC-SWMR-07}. 

Proof of the claim. 
Let us consider a correct process $p_j$ when it receives a message 
{\sc read}$()$ from $p_i$. It saves $w\_sync_i[i]$ in $sn$ and waits until
$w\_sync_j[i]\geq sn$ (lines~\ref{NC-SWMR-19}-\ref{NC-SWMR-20}). 
Due to Lemma~\ref{lemma-i-j-sn}, the predicate $w\_sync_j[i]\geq sn$ 
eventually becomes true at $p_j$. When this occurs, $p_j$ sends the message 
{\sc proceed}$()$ to $p_i$ (line~\ref{NC-SWMR-21}), which proves the claim.

Let us now consider the wait statement at line~\ref{NC-SWMR-09}, where 
$sn$ is the value of $w\_sync_i[i]$ when the  wait statement of 
line~\ref{NC-SWMR-07} terminates.
Let $p_j$ be a correct process. Due to Lemma~\ref{lemma-i-j-sn}
the predicate $w\_sync_i[j] \geq sn$ eventually holds.
 As this is true for any correct 
process $p_j$, $p_i$ eventually exits the wait statement, which concludes 
the proof of the lemma. 
\renewcommand{\toto}{lemma-read-termination}
\end{proofL}

\begin{lemma}
\label{lemma-read-atomicity}
The register that is built is atomic. 
\end{lemma}

\begin{proofL}
Let $read[i,x]$ be a read operation issued by a process $p_i$ which returns 
the value with sequence number $x$ (i.e., $history_i[x]$), 
and $write[y]$ be the write operation which writes the value with sequence 
number $y$  (i.e., $history_w[y]$). 
The proof of the  lemma is the consequence of the three following claims. 
\begin{itemize}
\vspace{-0.1cm}
\item Claim 1. 
If $read[i,x]$ terminates before $write[y]$ starts, then $x<y$.
\vspace{-0.2cm}
\item Claim 2. 
If $write[x]$ terminates before $read[i,y]$ starts, then $x\leq y$.
\vspace{-0.2cm}
\item Claim 3. 
If $read[i,x]$ terminates before $read[j,y]$ starts, then $x\leq y$.
\end{itemize}
Claim 1 states that no process can read from the future. 
Claim 2 states that no process can read overwritten values.
Claim 3 states that there is no new/old read inversion~\cite{AW04,R13-a}. \\

\noindent
Proof of Claim 1. \\ 
Due to Lemma~\ref{lemma-prefix-history}, the value returned by $read[i,x]$ 
is $history_i[x]= history_w[x]=v_x$. 
As each write generate a greater sequence number, 
and $p_w$ has not yet invoked ${\sf write}(v_y)$, we necessarily 
have $y>x$. \\

\noindent
Proof of Claim 2. \\ 
It follows from lines~\ref{NC-SWMR-01}-\ref{NC-SWMR-03} that when  
$write[x]$ terminates, there is a quorum $Q_w$ of at least $(n-t)$ processes 
$p_i$ such that $w\_sync_w[j]=x$. 
On another side, $read[i,y]$ obtains messages {\sc proceed}$()$ 
from a quorum $Q_r$ at least $(n-t)$ processes (lines~\ref{NC-SWMR-22}  
and~\ref{NC-SWMR-07}).  
As $|Q_w|\geq n-t$, $|Q_r|\geq n-t$, and $n-t>n/2$, 
we have $Q_w\cap Q_r\neq \emptyset$. Let $p_k$ be a process 
of $Q_w\cap Q_r$. As  $w\_sync_w[k]=x$,  and $w\_sync_k[k]\geq w\_sync_w[k]$
(Lemma~\ref{lemma-wsync-i-wrt-j}), and $write[x]$ is the last write before  
$read[i,y]$, we have $w\_sync_k[k]=x$ when  $read[i,y]$ starts. 

When $p_k$ received the message {\sc read}$()$ from $p_i$, we had
$w\_sync_k[k]=x$, and $p_k$ waited until $w\_sync_k[i]\geq x$
(line~\ref{NC-SWMR-20}) before sending the message {\sc proceed}$()$
that allowed $p_i$ to progress in its waiting at
line~\ref{NC-SWMR-07}.  As $w\_sync_i[i] \geq w\_sync_k[i]$
(Lemma~\ref{lemma-wsync-i-wrt-j}), it follows that we have
$w\_sync_i[i] \geq x$, when $p_i$ computes at line~\ref{NC-SWMR-08} the
sequence number $sn$ of the value it will return at
line~\ref{NC-SWMR-10}). Hence, the index $y=sn$ computed by $p_i$ 
at line~\ref{NC-SWMR-08} is such that $y=sn=w\_sync_i[i] \geq x$. ~\\

\noindent
Proof of Claim 3. \\ 
On one side, 
when $read[i,x]$ stops waiting at line~\ref{NC-SWMR-09}, there is a quorum 
$Q_{ri}$  of  at least $(n-t)$ processes $p_k$ such that $w\_sync_i[k] \geq x$
(predicate of line~\ref{NC-SWMR-09} at $p_i$).  
Due to Lemma~\ref{lemma-wsync-i-wrt-j}, we have then $w\_sync_k[k] \geq x$ 
for any process $p_k$ of $Q_{ri}$, when $read[i,x]$ terminates.

On the other side,  when $read[j,y]$ stops waiting at line~\ref{NC-SWMR-07} 
(which defines the value it returns, namely, $history_j[y]$),  there is
a quorum $Q_{rj}$ of  at least $(n-t)$ processes $p_\ell$  such that 
(due to the waiting predicate of line~\ref{NC-SWMR-20}) 
$w\_sync_\ell[j] \geq sn(\ell)$, where $sn(\ell)$ is the value of 
$w\_sync_\ell[\ell]$ when $p_\ell$ receives the message {\sc read}$()$
from $p_j$. 

As each of $Q_{ri}$ and $Q_{rj}$  contains at least $(n-t)$ processes, 
and there is a majority of correct processes, there is at least one correct 
process in their intersection, say $p_m$.
It follows that we have $w\_sync_m[m] \geq x$ when  $read[i,x]$ terminates, 
and  $w\_sync_m[j] \geq sn(m)$, where $sn(m)$ is the value of $w\_sync_m[m]$, 
when $p_m$ received the message {\sc read}$()$ from $p_j$. 
As  $w\_sync_m[m]$ never decreases, and $p_m$ receives the message 
{\sc read}$()$  from $p_j$ after  $read[i,x]$ terminated, 
we necessarily have  $sn(m)\geq x$. Hence,  $w\_sync_m[j] \geq x$, when 
$p_m$ sends {\sc proceed}$()$ to $p_j$. As (Lemma~\ref{lemma-wsync-i-wrt-j}) 
$w\_sync_j[j] \geq  w\_sync_m[j]$, it follows that the index $sn$ 
computed by $p_i$ at line~\ref{NC-SWMR-08} is such that $sn=y\geq x$.
\renewcommand{\toto}{lemma-read-atomicity}
\end{proofL}

\begin{theorem}
\label{theorem-main}
The algorithm described in Figure~{\em\ref{algo:two-bit-messages}} 
implements an {\em SWMR} atomic register in the system model 
${\cal CAMP}_{n,t}[t<n/2]$. 
\end{theorem}

\begin{proofT}
The theorem follows from  Lemma~\ref{lemma-write-termination}
and Lemma~\ref{lemma-read-termination} (Termination properties), and 
Lemma~\ref{lemma-read-atomicity} (Atomicity property). 
\renewcommand{\toto}{theorem-main}
\end{proofT}

\begin{theorem}
\label{theorem-property}
The algorithm described in Figure~{\em\ref{algo:two-bit-messages}} 
uses only four types of messages, and those carry no additional control 
information. Moreover, a read operation requires $O(n)$ messages, and a write
operation  requires $O(n^2)$ messages.  
\end{theorem}

\begin{proofT}
The  message content part of the theorem is trivial. 
A read generates $n$ messages {\sc read}$()$, and each of generates a message 
{\sc proceed}$()$. A write operation generates $(n-1)$ messages  
{\sc write}$(b,-)$ from the  writer to the other processes, and then each 
process forward once this message to each process. 
\renewcommand{\toto}{theorem-property}
\end{proofT}

\section{Concluding Remarks}
\label{sec:conclusion}

\paragraph{The aim and the paper}

As indicated in the introduction, our aim was to investigate the following 
question:  ``{\it How many bits of control information messages have to carry 
to implement an atomic register in ${\cal CAMP}_{n,t}[t<n/2]$?}''. 

As far as we know, all the previous works addressing this issue 
have reduced the size of control information with the use of  a ``modulo $n$''
implementation technique. 
Table~\ref{table-comparaison}  presents three algorithms plus ours. 
These three algorithms are the unbounded 
version of the ABD algorithm~\cite{ABD95}, its bounded version, and the bounded 
algorithm due to H. Attiya~\cite{A00}. They all associate a sequence number 
with each written value, but differently from ours, the last two  require  
each message to carry a ``modulo representative'' of a sequence number.

For each algorithm, the table  considers the number of messages 
it uses to implement the write operation  (line 1), the read operation
(line 2), the number of control bits carried by messages (line 3), 
the size of local memory used by each process (line 4), the time complexity 
of the write  operation (line 5), and the  time complexity of
the read operation (line 6), both  in a failure-free context. 
For time complexity it is assumed that message transfer delays are 
bounded  by $\Delta$, and local computations are instantaneous. 
The values appearing in the table for the bounded version of ABD and 
Attiya's algorithm are from~\cite{A00,R08}. 
The reader can see that the proposed algorithm is particularly efficient
from a time complexity point of view, namely, it is as good as the unbounded 
version of ABD. 

\begin{table}[ht]
\begin{center}
\renewcommand{\baselinestretch}{1}
\small
\begin{tabular}{c||c|c|c|c|c|}
\hline
line & What is & ABD95 \cite{ABD95} & ABD95 \cite{ABD95} &
                        H. Attiya's & Proposed  \\
number & measured & unbounded seq. nb  &  bounded seq. nb  & 
algorithm \cite{A00} & algorithm \\
\hline
1 &  \#msgs: write  & $O(n)$  &  $O(n^2)$  &  $O(n)$  &  $O(n^2)$  \\
\hline
2 &  \#msgs: read  & $O(n)$  &  $O(n^2)$  &   $O(n)$   &  $O(n)$ \\
\hline
3 & msg size  (bits) &  unbounded  &  $O(n^5)$  &   $O(n^3)$ & 2  \\
\hline
4 & local memory   &  unbounded    &  $O(n^6)$   &   $O(n^5)$ &  unbounded\\
\hline
5 & Time: write  & $2\Delta$ & $12\Delta$ & $14\Delta$ & $2\Delta$ \\
\hline
6 & Time: read   & $4\Delta$ & $12\Delta$ & $18\Delta$ & $ 4\Delta$\\
\hline 
\end{tabular}
\end{center}
\caption{A few algorithms implementing
an SWMR atomic register in ${\cal CAMP}_{n,t}[t<n/2]$}
\label{table-comparaison} 
\end{table}

\paragraph{The result presented in  the paper}
As we have seen, our algorithm also uses sequence numbers, but those 
remain local. Only four types of messages are used, which  means 
that each implementation message carries only two bits of control information.
Moreover, only two message types carry a data value,  the other two carry 
no data at all. Hence,  this paper answers a long lasting question: ``{\it
it is possible to implement an atomic register, despite asynchrony and crashes 
of a minority of processes, with messages whose control part is constant?}''. 

The unbounded feature of the proposed algorithm (when looking at the local
memory size)  is due to the fact that the algorithm 
introduces a fault-tolerant version of  a ``synchronizer''\footnote{As
 introduced in~\cite{A85}, and presented in textbooks such
as~\cite{AW04,L96,R13}.} suited to the implementation of an atomic register, 
which  disseminates  new values, each  traveling between  
each pair of processes in both directions, in such a way that a strong 
synchronization is ensured between any pair of processes, independently from
the other processes, (namely, 
$\forall i,j:~ 0\leq |w\_sync_i[j] - w\_sync_j[i]| \leq 1$).
This fault-tolerant 
synchronization is strong enough to allow sequence numbers to be eliminated 
from messages. Unfortunately, it does not seem  appropriate to allow a local
modulo-based representation of sequence numbers at each process.

In addition to its theoretical interest, and thanks to its time complexity,  
the proposed algorithm is also interesting from a practical point of view. 
Due to the  $O(n)$  message cost of its read operation, it can benefit to 
read-dominated applications and, more generally, to any setting where the 
communication cost (time and message size) is the critical parameter\footnote{
In addition to the way they use sequence numbers, 
an  interesting design difference between our algorithm and ABD-like 
algorithms is the following. When a process receives a message {\sc read}$()$, 
it has two possibilities. Either send by return the last written value it
knows, as done in ABD-like algorithms. Or wait until it knows that the 
sender has  a value as up to date as it  own value, and only then send it
 a signal, as done in our algorithm with the message {\sc proceed}$()$.}. 

\paragraph{A problem that remains open}
According to the previous discussion, a problem that still remains open 
is the following. Is it  possible  to design an implementation 
where (a) a constant number of bits is sufficient to encode 
the control information carried by messages, and   
(b) the sequence numbers have a local modulo-based implementation?
We are inclined to think that this is not possible. 

\Xomit{
\begin{itemize}
\vspace{-0.2cm}
\item ABD95  with unbounded seq nbs: \\
- each operation requires $O(n)$ messages,\\
- unbounded local memory, and unbounded messages,\\
- Read takes $2\Delta$, Write  takes $4\Delta$.
\vspace{-0.2cm}
\item ABD95  with bounded seq nbs: \\
- each operation requires $O(n^2)$ messages,\\
- bounded local memory: $O(n^6)$ bits, \\
- each message:  $O(n^5)$ bits,\\
- Read and write take $12\Delta$.
\vspace{-0.2cm}
\item A00  with bounded seq nbs: \\
- each operation requires $O(n)$ messages,\\
- each message requires $O(n^3)$ bits,\\
- read takes up to $18\Delta$, and write takes $14\Delta$,\\
- the writer uses  $O(n^5)$ bits of local memory, and the reader uses 
 $O(n^4)$ bits.
\vspace{-0.2cm}
\item 
Proposed algorithm\\
- read  requires $O(n)$ messages, write requires  $O(n^2)$  messages (?),\\
- message size: 2 bits  (three message types),\\
- read takes $2\Delta$ in ``good'' circumstances, 
  and at most $3\Delta$ or $4\Delta$ in ``bad'' circumstances,\\
- write takes $2\Delta$,\\
- unbounded local memory.
\end{itemize}
}

\section*{Acknowledgments}
This work has been partially supported by the  French  ANR  project DISPLEXITY, 
which is devoted to  computability and  complexity in distributed computing, 
and the Franco-German ANR project DISCMAT devoted to connections between 
mathematics  and distributed computing.


\end{document}